Carrier transport in reverse-biased graphene/semiconductor Schottky junctions

D. Tomer*[1], S. Rajput[1], L. J. Hudy[1], C. H. Li[2], and L. Li[1]

[1]Department of Physics, University of Wisconsin, Milwaukee, WI, 53211
[2]Naval Research Laboratory, Washington, DC 20375, USA

Reverse-biased graphene (Gr)/semiconductor Schottky diodes exhibit much enhanced sensitivity for gas sensing. However, carrier transport across the junctions is not fully understood yet. Here, Gr/SiC, Gr/GaAs and Gr/Si Schottky junctions under reverse-bias are investigated by temperature-dependent current-voltage measurements. A reduction in barrier height with increasing reverse-bias is observed for all junctions, suggesting electric-field enhanced thermionic emission. Further analysis of the field dependence of the reverse current reveals that while carrier transport in Gr/SiC Schottky junctions follows the Poole-Frenkel mechanism, it deviates from both the Poole-Frankel and Schottky mechanisms in Gr/Si and Gr/GaAs junctions, particularly for low temperatures and fields.

*Correspondence and requests for materials should be addressed to D.T. (dtomer@uwm.edu)

The high electrical conductivity, tunable Fermi level, and high surface to volume ratio of graphene (Gr) make it a promising material for optoelectronic and sensing applications[1-3]. In particular, when graphene is interfaced with semiconductors it forms Schottky junctions[4] that can be used in solar cells, photo-detectors, three-terminal transistors, and gas sensors[5-11]. Earlier work has shown that these Schottky diodes under forward-bias can be generally understood within the thermionic emission (TE) model[4,12-15]. Recent work has further indicated that the non-ideal behavior of the temperature dependent barrier height and ideality factor greater than one can be attributed to spatial inhomogeneities commonly observed at the junction interface[16,17]. A modified TE model assuming a Gaussian distribution of the barrier height offers a better account of carrier transport in these graphene/semiconductor Schottky junctions[16,17]. For near ideal Gr/Si junction, Landauer transport mechanism is suggested[18]. Under reverse-bias, on the other hand, these Schottky diodes can exhibit much improved performance for gas sensing compared to field effect devices, primarily due to the exponential dependence of the reverse current on the Schottky barrier height (SBH)[9]. However, few studies have addressed carrier transport under reverse-bias in Gr/semiconductor Schottky junctions[18].

In this work, we fabricate Gr/semiconductor (SiC, GaAs and Si) Schottky junctions by transferring chemical vapor deposited monolayer graphene onto hydrogen-terminated hexagonal SiC (both the (0001) Si-face and (000$\bar{1}$) C-face) and Si (111), and sulfur-terminated GaAs (100) substrates. Details of device fabrication and forward-bias characteristics of these Schottky diodes can be found elsewhere[16,17]. Here, we present studies of the temperature- and electric- field dependence of the junction current under reverse-bias, where we find that while carrier transport in reverse-biased Gr/SiC Schottky junctions follows the Poole-Frenkel mechanism, it deviates from both the Poole-Frankel and Schottky mechanisms in Gr/Si and Gr/GaAs junctions, particularly for low temperatures and electric fields.

All junctions exhibit rectifying behaviors, as shown for example in Fig. 1(a) for Gr/GaAs at 310 K[16,17]. Under the reverse-bias, however, the current rises with increasing bias voltage, as better seen in the semi-logarithmic plots for Gr/GaAs and Si-face SiC diodes between 250 and 340 K (Fig. (b) & (c)). This is clearly inconsistent with the simple thermionic emission picture[19], where for V<-3$k_B$T/q ($k_B$ is the Plank constant and $q$ the charge), the reserve bias current should saturate:

$$I_r(T) = AA^*T^2 \exp\left(\frac{-q\phi_{bo}}{k_B T}\right) \tag{1}$$

with $A$ the junction contact area, $A^*$ the effective Richardson constant (1.46x10$^6$, 1.12x10$^6$, and 0.41x10$^4$ Am$^{-2}$K$^{-2}$ for SiC, Si and GaAs, respectively), and $\phi_{bo}$ the zero bias barrier height. Similar behaviors are also observed for Gr/C-SiC and Gr/Si Schottky junctions.

The non-saturating current under reverse-bias suggests that the barrier height is a function of the bias voltage, as calculated following Eq. (1) and shown in Fig. 2. At 310 K. The calculated $\phi_{bo}$ decreases with increasing reverse-bias for all junctions, similar to a previous work on Gr/Si Schottky junctions[18], with the Gr/GaAs junction showing a lower barrier and larger variation. These behaviors suggest low interfacial states at these graphene Schottky junctions, since they are known to pin the Fermi level in the semiconductors, which leads to reverse current saturation in conventional metal/semiconductor junctions[20].

To account for the reduction of barrier height with increasing reverse-bias, electric-field enhanced thermionic emission is further investigated following the Poole-Frenkel[21] and Schottky[22] mechanisms. The reverse current considering Poole-Frenkel emission is given by:[21]

$$I_r \propto E \exp(\frac{q}{k_B T}\sqrt{\frac{qE}{\pi \varepsilon_s}}) \qquad (2)$$

Whereas in the case of Schottky emission it is given by:[21]

$$I_r \propto T^2 \exp(\frac{q}{2k_B T}\sqrt{\frac{qE}{\pi \varepsilon_s}}) \qquad (3)$$

Here $E$ is the applied electric field, calculated by $E = \sqrt{\frac{2qN_D}{\varepsilon_s}(V + V_{bi} - \frac{k_B T}{q})}$, where $\varepsilon_s$ is the relative dielectric constant of the semiconductor (~9.66 for SiC[23]), $N_D$ the donor density of the semiconductor (~10$^{18}$ cm$^{-3}$)[16], $V$ applied bias, and $V_{bi}$ the built-in potential. The built-in potential is a function of forward-biased SBH and the effective density of states in semiconductor conduction band ($N_C$) at room temperature[19], taken as 1.69x10$^{19}$ for Si-SiC[23]. The mean SBH values of 1.16 eV are taken for Gr/Si-SiC Schottky junction[16].

Thus, if the Poole-Frenkel effect contributes to the reverse current, then the plot of *ln(I$_r$/E)* versus *E$^{1/2}$* should be linear. Similarly if a linear plot is found for *ln(I$_r$/T$^2$)* versus *E$^{1/2}$*, then the Schottky mechanism is present. Figures 3(a) and (b) shows plots of *ln(I$_r$/E)* and *ln(I$_r$/T$^2$)* as a function of *E$^{1/2}$*, respectively, for the Gr/Si-SiC Schottky junction. Clearly, both are near linear for all temperatures, indicating that both Schottky and Poole-Frenkel emissions are present.

To distinguish which mechanism the carrier transport is dominated by, we calculate the emission coefficient following[21]

$$S = \frac{q}{nkT}\sqrt{\frac{q}{\pi\varepsilon_s}} \qquad (4)$$

where *n=1* for Poole-Frenkel emission and *n=2* for Schottky emission. The calculated coefficients are compared with that obtained by curve fitting for both Poole-Frenkel and Schottky emissions at different temperatures, as shown in Table *1* between 250 and 340 K for the Gr/Si-SiC Schottky junction. For the Poole-Frenkel emission, the experimental values are almost ~2-2.3 times of the calculated at all temperatures. For Schottky emission, the experimental values are ~3.5 times larger. Similar trend is also found for Gr/C-SiC junction. Thus, carrier transport in graphene/SiC Schottky junctions under reverse-bias is more consistent with the Poole-Frenkel mechanism for these temperatures.

Similar analysis was performed for Gr/GaAs and Gr/Si junctions, as shown in Fig. 4, where $\varepsilon_s$ is taken as $12.9^{24}$ and $11.7^{25}$, $N_D \sim 10^{16}$ and $10^{17}$ cm$^{-3}$, $N_C = 4.70 \times 10^{17}$ and $2.86 \times 10^{19}$ cm$^{-3}$ for GaAs, and Si, respectively[23], and the mean SBH values of 1.14 and 0.76 eV are taken for Gr/Si and Gr/GaAs Schottky junctions[17]. For Gr/GaAs (Fig. 4(a&b), both the Poole-Frenkel and Schottky emission plots are linear above 310 K, suggesting a possible co-existence of both mechanisms at higher temperatures. Again the Poole-Frenkel and Schottky emission coefficients must be considered to identify their contributions to carrier transport. At 340 K, the Poole-Frenkel coefficient obtained from the fit is 0.119 (V/cm)$^{1/2}$, ~ 16 times the calculated value of 0.007 (V/cm)$^{1/2}$. On the other hand for the Schottky emission, the experimental value (0.128 (V/cm)$^{1/2}$) is ~35 times greater than that calculated (0.0036 (V/cm)$^{1/2}$).

For the Gr/Si Schottky junction, the Poole-Frenkel and Schottky emission plots are shown in Fig. 4 (c) & (d). In both cases, although they are still linear at 340 K, non-linearity is found for temperatures below 310 K, particularly in the low electric field region. Fittings to the linear plots at 340 K yield an experimental emission coefficient of 0.0615 $(V/cm)^{1/2}$, ~ 8 times greater than the calculated value (0.0076 $(V/cm)^{1/2}$) for the Poole-Frenkel mechanism. For the Schottky emission, the experimental value (0.0658 $(V/cm)^{1/2}$) is ~17 times greater than that calculated (0.0038 $(V/cm)^{1/2}$).

These results suggest that carrier transport in the reverse-biased Gr/GaAs and Gr/Si Schottky diodes deviate from the Poole-Frenkel and Schottky emission, particularly at low temperatures and electric fields. This may be due the much larger depletion width, ~ 0.1-0.5 μm, for the Gr/Si and Gr/GaAs Schottky junctions, comparable to that reported for Gr/Si[26]. For comparison, the depletion width is only ~ 35-48 nm for the Gr/Si-SiC junctions. In addition, other conduction mechanisms such as bias dependent doping[18], i.e., electric field dependence of the Fermi level in graphene, should also be taken into account in these non-linear regions.

In conclusion, reverse-biased Gr/SiC, Gr/GaAs, and Gr/Si Schottky junctions are studied by temperature dependent I-V measurements between 250 and 340 K. A reduction in barrier height with increasing reverse-bias is observed for all junctions, consistent with electric-field enhanced thermionic emission. Analysis of the field dependence of the reverse current reveals that while carrier transport in Gr/SiC Schottky junctions follows the Poole-Frenkel mechanism, it deviates from both the Poole-Frankel and Schottky mechanisms in Gr/Si and Gr/GaAs junctions, particularly in the low temperature and field regimes, where field dependent doping in graphene should also be taken into account. Our findings present the first evidence for electric field enhanced thermionic emission in graphene/semiconductor Schottky junctions under reverse bias, providing insights on the carrier transport mechanisms that are critical for improving functions of graphene-based devices.

**Acknowledgment**

Research supported by the U.S. Department of Energy, Office of Basic Energy Sciences, Division of Materials Sciences and Engineering under Award DE-FG02-07ER46228.

**Figure captions**

**Figure 1** (color online) (a) I-V curve of Gr/GaAs Schottky junction at 310K showing a rectifying behavior (inset: schematic diagram of the devices). Temperature dependent I-V characteristics of (b) Gr/Si-SiC and (c) Gr/GaAs Schottky junctions in the reverse-bias region between 250 and 340 K

**Figure 2** (color online) Calculated Schottky barrier height $\phi_{b0}$ as a function of reverse-bias voltage at 310 K.

**Figure 3** (color online) Temperature dependent (a) Poole-Frenkel and (b) Schottky emission plots for Gr/Si-SiC Schottky junctions.

**Figure 4** (color online) Temperature dependent (a,c) Poole-Frenkel and (b,d) Schottky emission plots for Gr/GaAs and Gr/Si Schottky junctions.

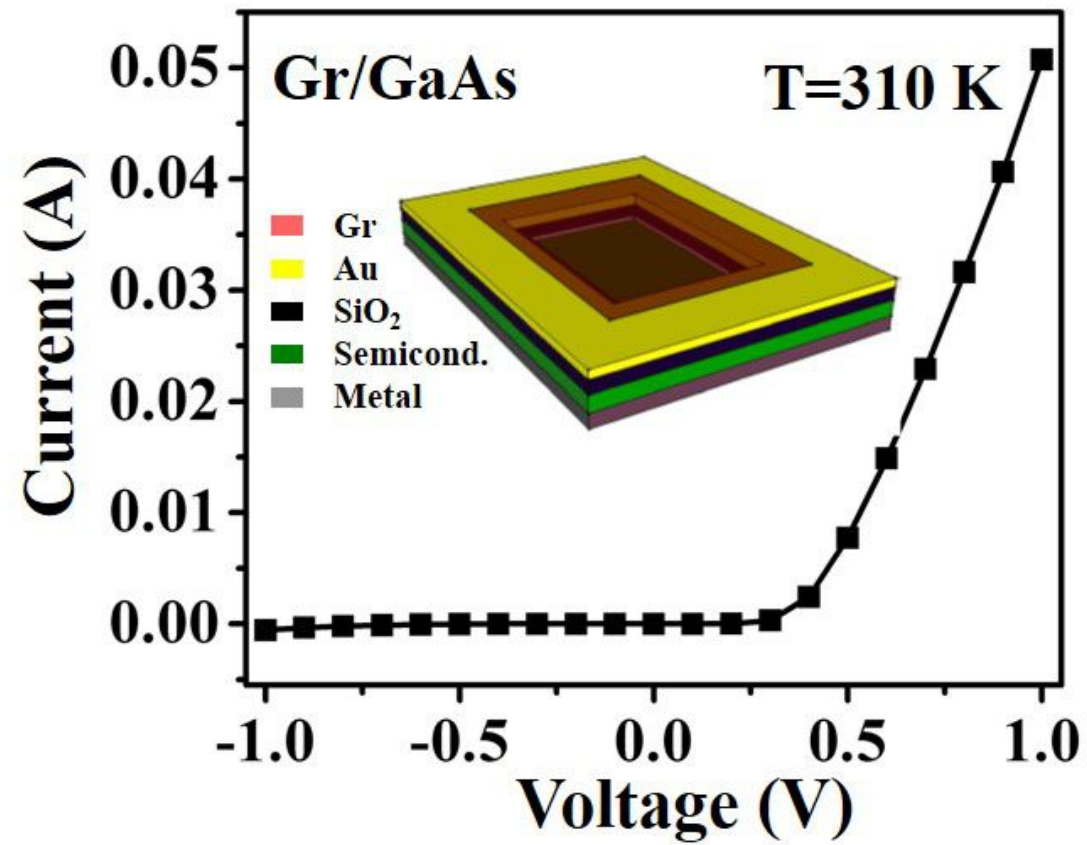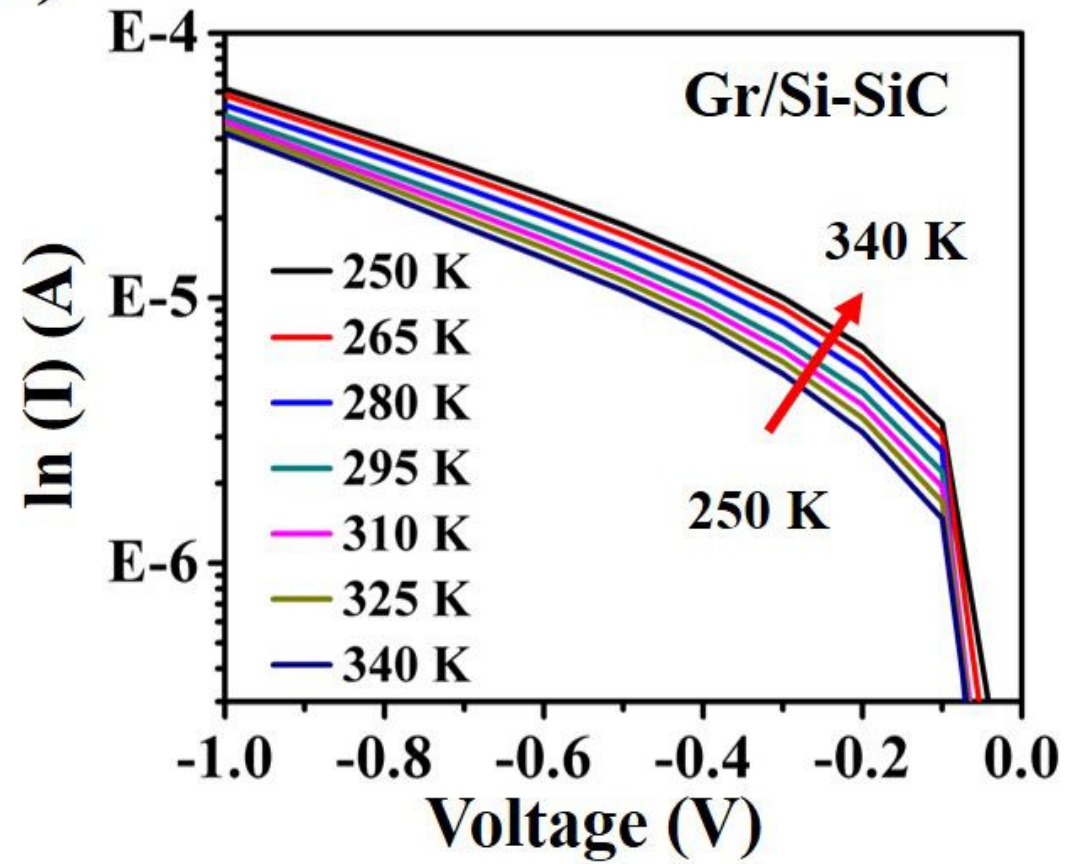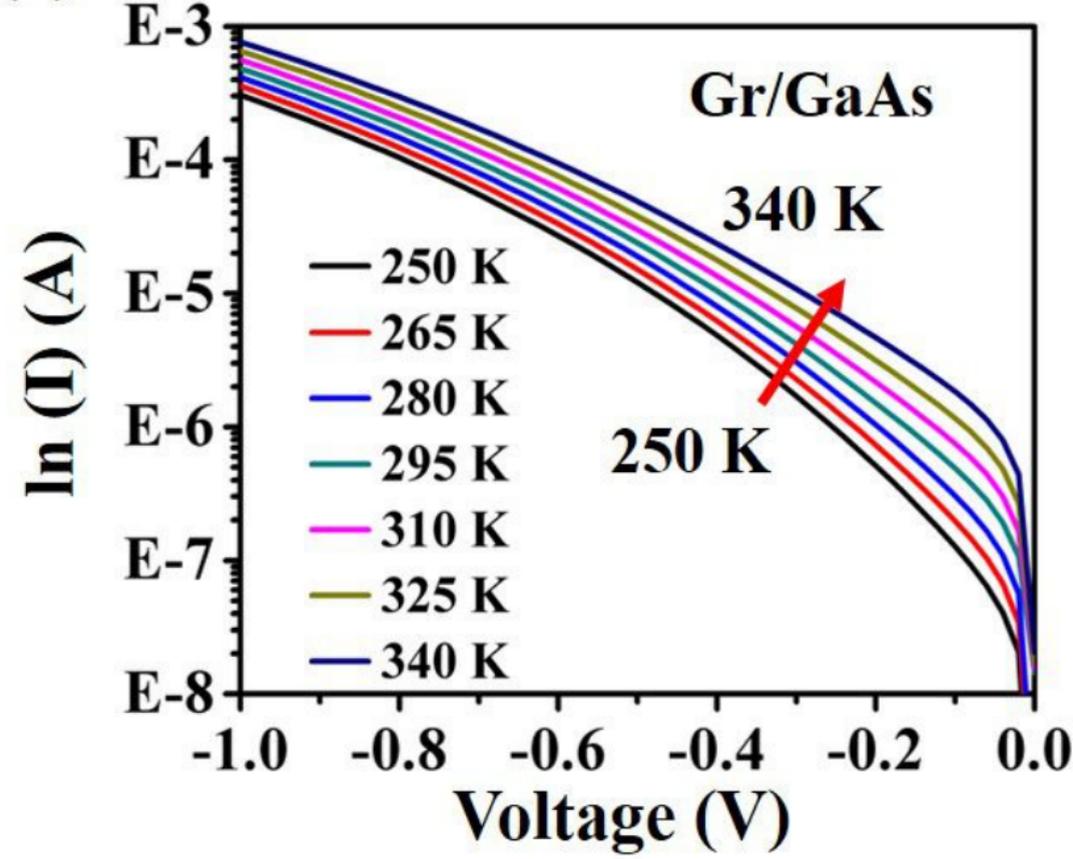

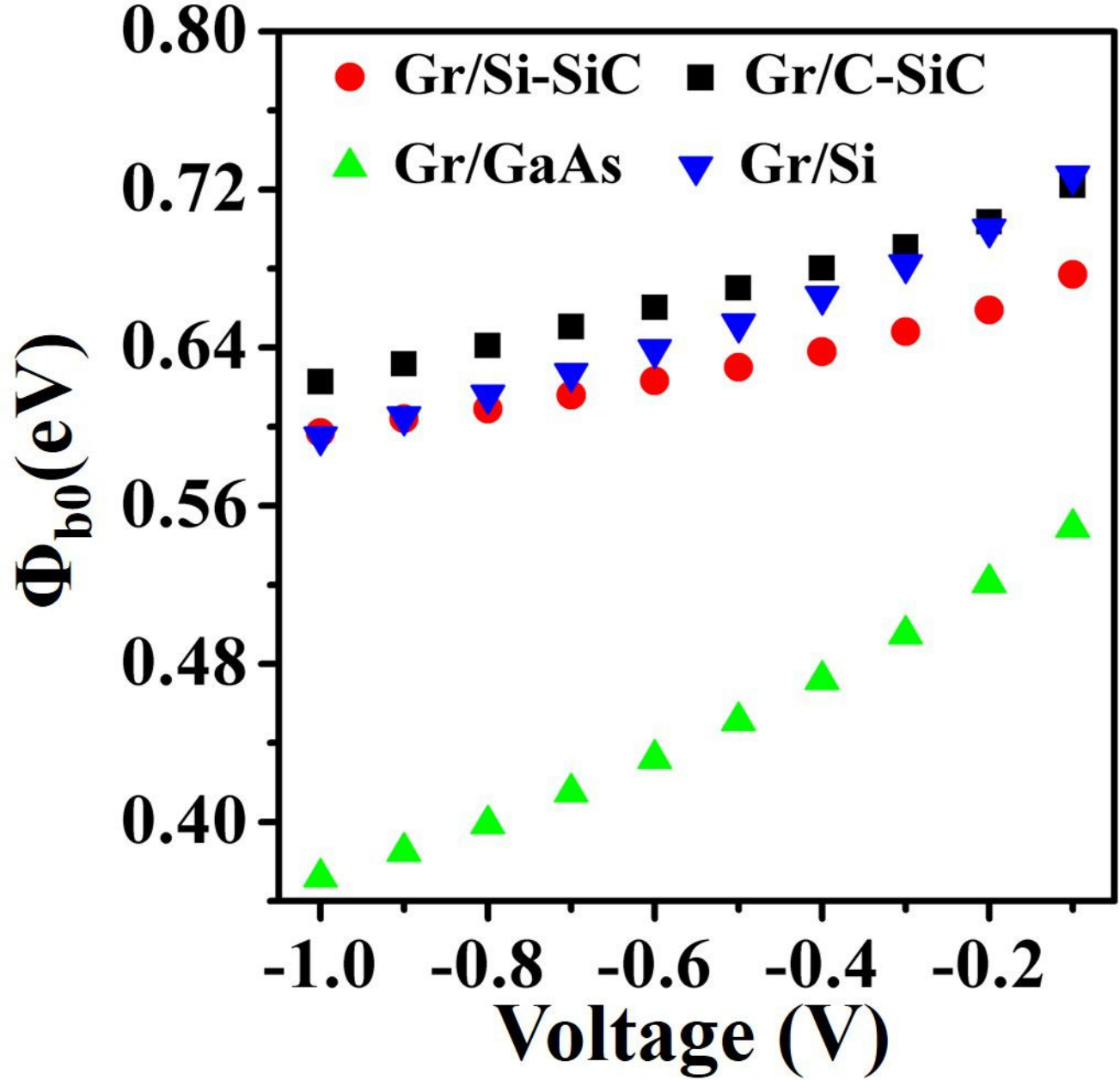

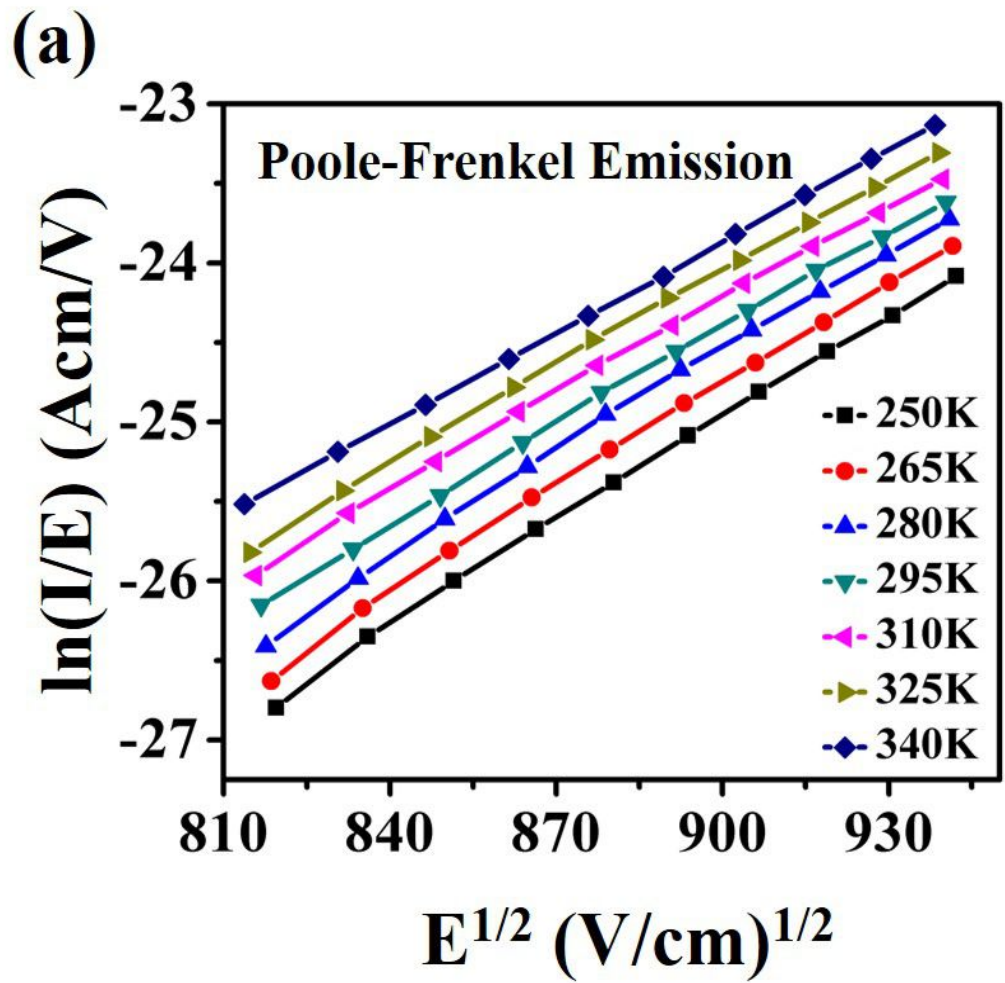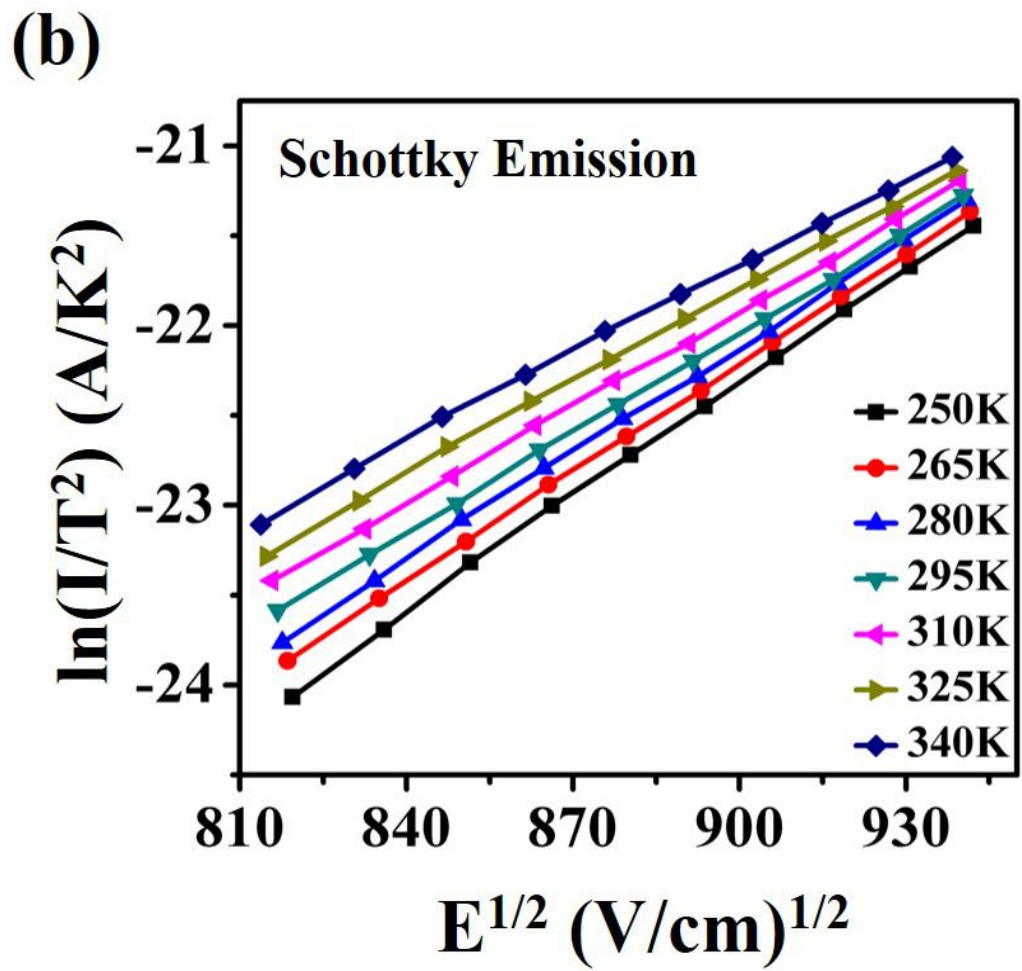

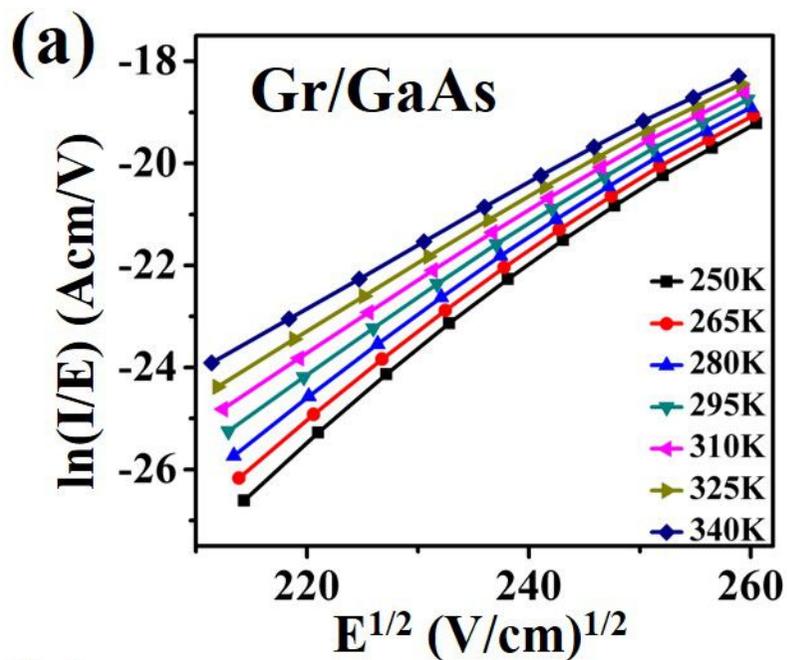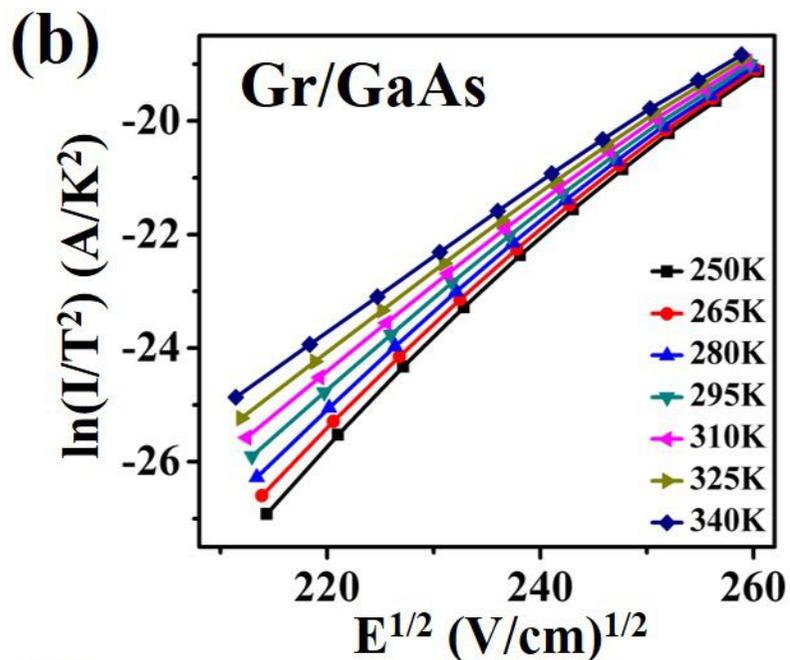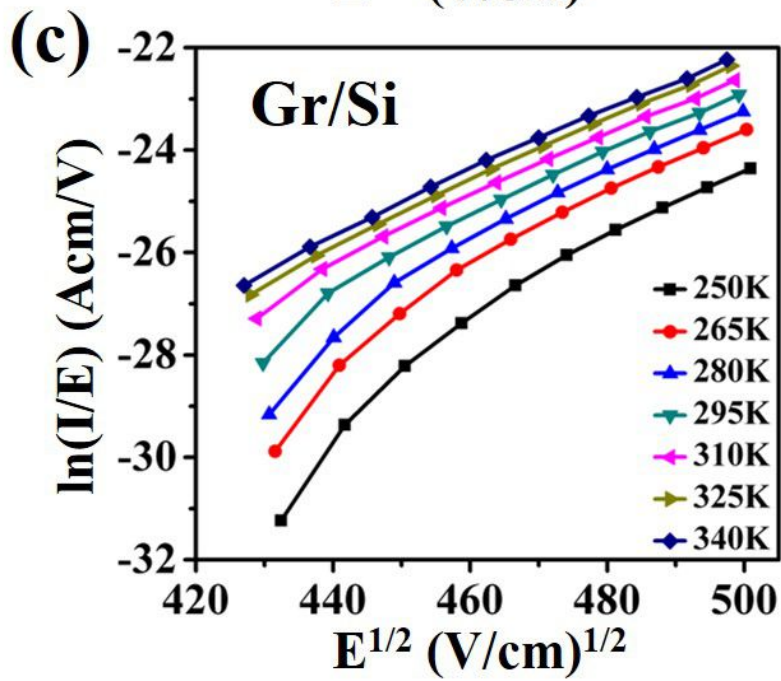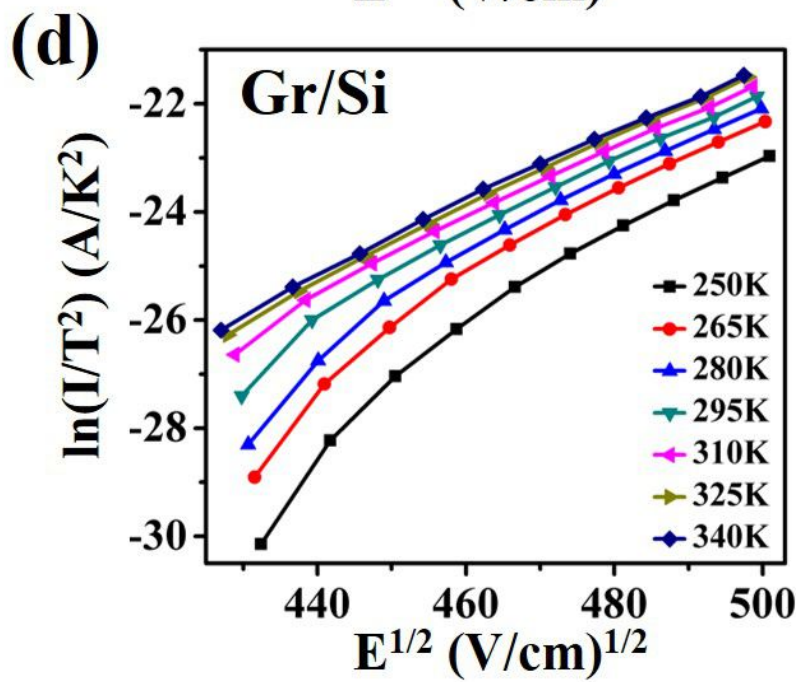

**Table 1** Comparison of calculated and experimental Poole-Frenkel and Schottky emission coefficients between 250 and 340 K for the Gr/Si-SiC Schottky junction

| Temperature (K) | Poole-Frenkel emission (V/cm)$^{1/2}$ | | Schottky emission (V/cm)$^{1/2}$ | |
|---|---|---|---|---|
| | Calculated | From fit | Calculated | From fit |
| 250 | 0.0113 | 0.0224 | 0.0056 | 0.0213 |
| 265 | 0.0107 | 0.0219 | 0.0053 | 0.0202 |
| 280 | 0.0101 | 0.0215 | 0.0050 | 0.0198 |
| 295 | 0.0096 | 0.0206 | 0.0048 | 0.0185 |
| 310 | 0.0091 | 0.0199 | 0.0046 | 0.0178 |
| 325 | 0.0087 | 0.0194 | 0.0044 | 0.0171 |
| 340 | 0.0083 | 0.0191 | 0.0041 | 0.0161 |